\documentclass{article}

\usepackage{amsmath}
\usepackage{amssymb}

\usepackage{cite}

\usepackage{geometry}

\newcommand{\vk}{\varkappa}

\newcommand{\im}{\mathrm{Im}}
\newcommand{\re}{\mathrm{Re}}
\newcommand{\dn}{\mathrm{dn}}

\allowdisplaybreaks

\title{Spectral curves for the rogue waves }

\author{Aleksandr O. Smirnov, Vladimir B. Matveev, Yury A. Gusman and Nikita V. Landa}

\begin{document}

\maketitle

\begin{abstract}
 Here we find the spectral curves, corresponding  to the  known rational or  quasi-rational solutions of  AKNS hierarchy equations, ultimately connected with the modeling of the rogue waves events in the optical waveguides and in hydrodynamics.  We also determine  spectral curves for the multi-phase  trigonometric, hyperbolic and elliptic solutions for the same hierarchy.
It seams  that the nature of the related spectral curves was not sufficiently discussed  in existing
  literature.


\end{abstract}

\section*{Introduction}

The quasi-rational multi-parametric solutions for the NLS equation, explaining a phenomena of the multiple rogue waves generation, were first obtained in \cite{DGKM10}  in 2010, (see also   \cite{DubMat11}, and  \cite{DubMat13}), - using a slightly modified technique of work \cite{EKKe}.  Recently this approach was extended by two of the authors  in \cite{Sm15a, Sm16tmfe} to the whole AKNS hierarchy.

It was mentioned  in \cite{DGKM10}  that it is possible to attend the same goal  by using a properly generalized Darboux transformation method in spirit of the works \cite{Mat92a} and \cite{Mat92b}, - to the focusing NLS equation. This was  done in  \cite{GenDT11}.  See also \cite{AKA2} where an iterative  application of onset Darboux transformation was used.

At the same time,  Hirota method was successfully applied in \cite{Ohta12} to the same problem. Let us emphasize that  three aforementioned approaches were developed \textbf{without any use of the spectral curves}, associated  with  the finite gap solutions of the NLS  equation.

The 4-th approach,  also suggested in \cite{DGKM10},  is to start from  the finite-gap solutions of the focusing NLS equation
associated with nonsingular hyperelliptic curves,  and to consider an appropriate passage to the limit, corresponding to the confluence of several branch points of the spectral curve.  To some extent, this later approach was first considered  in \cite{IRSe, IAKe},  resulting in explicit description of various kinds of modulation instability  for the focusing NLS model, described by means of elliptic or trigonometric functions.
To obtain  the aforementioned quasi-rational solutions it was necessary to go from the results of \cite{IRSe}  to  a further  degeneration of  spectral curves and  related solutions,  taking care to keep a maximal number of free parameters at the end.  This was realized  in a number of works  by P.Gaillard, (see for instance \cite{GP11, GP13, GP13a}, see also further comments about theses works in \cite{DubMat13}).

Here in this work  we explain,  how to solve the inverse problem, - i.e. how, for various kinds of explicit solutions,  obtained without any apriory contact with  spectral curves, - to restore these spectral curves.   Thus, using quite elementary methods, valid for the whole AKNS hierarchy. we reply to the frequently posed question:   what spectral curves corresponds to rank $n$ multiple rogue waves solutions, or higher Peregrin breathers, or  trigonometric breathers and multi-breathers solutions?

Assuming the solutions are expressed  by means  of trigonometric, hyperbolic or rational functions,- we show that  the related spectral curves have multiple branch points. The goal of the work is to derive and analyze  equations of the spectral curves for known solutions, finding  the relations  between  parameters of solutions and  parameters of  the spectral curves.
This will allow us to  carry out the degeneration process of generic  algebro-geometric multi-phase solutions to solutions, expressed by means of elementary or elliptic  functions and  investigate new solutions of the AKNS hierarchy equation.
It will also help to establish   correspondence between  parametrizations of the multi-phase solutions
of AKNS hierarchy equations, obtained by different methods.

\section{Brief introduction to AKNS hierarchy}

For further details see \cite{AKNS}  and especially \cite{Sm16tmfe, Sm15a} in our context.
AKNS hierarchy equations follow from compatibility conditions  of the systems:
\begin{equation*}
\begin{cases}
\Psi_x=U \Psi,\\
\Psi_{t_k}=V_k \Psi,
\end{cases}
\end{equation*}

\begin{gather*}
U:=\lambda J+U^0,\quad V_1:=2\lambda U +V_{1}^0,\quad V_{k+1}:=2\lambda V_k+V_{k+1}^0,\quad k\geqslant1\\
J:=\begin{pmatrix} -i&0\\0&i \end{pmatrix}, \quad
U^0:=\begin{pmatrix} 0&ip\\ -iq&0 \end{pmatrix}.
\end{gather*}
The conditions $(\Psi_x)_{t_k}=(\Psi_{t_k})_x$ implies $V_k$, recursive
 relations for off-diagonal elements of matrices $V_k^0$, and the relations between diagonal and off-diagonal elements of these matrices.
\begin{equation*}
[J,V_1^0]=2(U ^0)_x,\quad [J,V_{k+1}^0]=2(V_k^0)_x+2[V_k^0,U ^0],\quad k\geqslant 1.
\end{equation*}
In particular,
\begin{equation*}
V_1^0=\begin{pmatrix}-ipq& -p_x\\-q_x &ipq\end{pmatrix},\quad
V_2^0=\begin{pmatrix} p_xq-q_xp& 2ip^2q-ip_{xx}\\
-2iq^2p+iq_{xx}& q_xp-p_xq\end{pmatrix}.
\end{equation*}

The integrable nonlinear equations from AKNS hierarchy are derived from the recursion relations
\begin{equation} \label{eq:Ut}
(U^0)_{t_k}=(V_k^0)_x+[V_k^0,U^0]=\dfrac12[J,V_{k+1}^0]. \tag{*}
\end{equation}

Let us list first five members of AKNS hierarchy and their  reduced versions, starting the RAKNS hierarchy:
\begin{equation}
\begin{cases}
ip_{t_1}+p_{xx} - 2p^2q=0,\\
-iq_{t_1}+q_{xx} - 2q^2p=0.
\end{cases} \label{eq:coup.nls}
\end{equation}
When $q=\mp p^\ast$ it  reduces to  a focusing nonlinear Schr\"odinger equation (NLS):
\begin{equation} \label{eq:nls}
ip_{t_1}+p_{xx}+2|p|^2p=0, \tag{\ref{eq:coup.nls}a}
\end{equation}
or to its defocusing version:
\begin{equation*}
ip_{t_1}+p_{xx} - 2|p|^2p=0,
\end{equation*}

Next member of AKNS hierarchy is called  coupled modified Korteweg-de Vries (cmKdV) system:
\begin{equation}
\begin{cases}
p_{t_2}+p_{xxx}-6pqp_x=0,\\
q_{t_2}+q_{xxx}-6pqq_x=0.
\end{cases} \label{eq:coup.mkdv}
\end{equation}
Setting  $q = - p^\ast$  it  reduces to  complex    modified Korteweg de Vries  (cmKdV) equation:
\begin{equation} \label{eq:cmkdv}
p_{t_2}+p_{xxx}+6|p|^2p_x=0. \tag{\ref{eq:coup.mkdv}a}
\end{equation}
Taking $q=\pm p$ in coupled modified Korteweg-de Vries system, we get usual (real) mKdV equation:
\begin{equation*}
p_{t_2}+p_{xxx} \pm 6p^2p_x=0. \label{eq:mkdv}
\end{equation*}

The third member of   AKNS hierarchy:
\begin{equation}
\begin{cases}
ip_{t_3}-p_{xxxx}+8pqp_{xx}+2p^2q_{xx}+6p_x^2q+4pp_xq_x-6p^3q^2=0,\\
-iq_{t_3}-q_{xxxx}+8pqq_{xx}+2q^2p_{xx}+6pq_x^2+4qp_xq_x-6p^2q^3=0.
\end{cases} \label{eq:coup.akns3}
\end{equation}
Under the constraints $q=-p^\ast$, $t_3=-t$ it  reduces \eqref{eq:coup.akns3} to
 Lakshmanan-Porsezian-Daniel (LPD) equation  \cite{GNLS88,GNLS92,GNLS93}:
\begin{equation} \label{eq:lpd}
ip_{t}+p_{xxxx}+8|p|^2p_{xx}+2p^2p^{\ast}_{xx}+6p_x^2p^{\ast}+4p|p_x|^2+6|p|^4p=0. \tag{\ref{eq:coup.akns3}a}
\end{equation}

The fourth member  of AKNS  hierarchy:
\begin{equation}
\begin{aligned}
&p_{t_4}-p_{5x}+10pqp_{xxx}+20qp_{x}p_{xx}+10pq_{x}p_{xx}+10pp_{x}q_{xx}+10p_{x}^2q_{x}-30q^2p^2p_{x}=0,\\
&q_{t_4}-q_{5x}+10qpq_{xxx}+20pq_{x}q_{xx}+10qp_{x}q_{xx}+10qq_{x}p_{xx}+10q_{x}^2p_{x}-30q^2p^2q_{x}=0
\end{aligned} \label{eq:coup.akns4}
\end{equation}

Taking  $q=-p^\ast$ and $t_4=-t$,  it reduces to:
\begin{equation} \label{eq:akns4}
p_{t}+p_{5x}+10|p|^2p_{xxx}+20p_{xx}p_xp^\ast+10(|p_x|^2p)_x+30|p|^4p_x=0. \tag{\ref{eq:coup.akns4}a}
\end{equation}
Surprisingly as short as  \eqref{eq:lpd}.

The fifth member  of AKNS hierarchy:
\begin{multline*}
ip_{t_5}+p_{6x}-12pqp_{xxxx}-2p^2q_{xxxx}-30p_{x}qp_{xxx}-18pq_{x}p_{xxx}-8pp_{x}q_{xxx}-\\
-50p_{x}q_{x}p_{xx}+50q^2p^2p_{xx}-20p_{xx}^2q-22pq_{xx}p_{xx}-20q_{xx}p_{x}^2+20q_{xx}qp^3+\\
+10p^3q_{x}^2+70q^2pp_{x}^2+60qp^2p_{x}q_{x}-20q^3p^4=0,
\end{multline*}
\begin{multline*}
-iq_{t_5}+q_{6x}-12qpq_{xxxx}-2q^2p_{xxxx}-30q_{x}pq_{xxx}-18qp_{x}q_{xxx}-8qq_{x}p_{xxx}-\\
-50q_{x}p_{x}q_{xx}+50q^2p^2q_{xx}-20q_{xx}^2p-22qp_{xx}q_{xx}-20p_{xx}q_{x}^2+20p_{xx}q^3p+\\
+10q^3p_{x}^2+70qp^2q_{x}^2+60q^2pq_{x}p_{x}-20q^4p^3=0.
\end{multline*}
For $q=-p^\ast$, it reduces  to a single scalar equation: \\
\begin{multline*} \label{eq:akns5}
ip_{t_5}+p_{6x}+12|p|^2p_{xxxx}+2p^2p^\ast_{xxxx}+30p_{xxx}p_xp^\ast+18p_{xxx}pp^\ast_x+8p_xpp^\ast_{xxx}+\\
+50p_{xx}|p_x|^2+50p_{xx}|p|^4+20p_{xx}^2p^\ast+22|p_{xx}|^2p+20p_x^2p^\ast_{xx}+20|p|^2p^2p^\ast_{xx}+\\
+10p^3(p^\ast_x)^2+70p_x^2|p|^2p^\ast+60|p|^2|p_x|^2p+20|p|^6p =0.
\end{multline*}

The system,  corresponding to the $k$-th member of AKNS hierarchy, can be written as follows:
\begin{equation*}
\begin{cases}
p_{t_k}=i^kH_k(p,q),\\
q_{t_k}=(-i)^kH_k(q,p).
\end{cases}
\end{equation*}
Explicit formulas  for the functions $H_k(p,q)$  for $k= 1, 2,\ldots 5$  are given above.  For higher values of $k$  they can be easily obtained  by use of the symbolic computation from \eqref{eq:Ut}, and for $k=6,7$ they can be found in \cite{AKACB} where $K_j(x,t)=H_{j+1}(p,-p^\ast)$.
All  equations  of the reduced AKNS hierarchy (RAKNS hierarchy) and, in particular,   \eqref{eq:nls}, \eqref{eq:cmkdv}, \eqref{eq:lpd}, \eqref{eq:akns4} \ etc. can be
written as
\begin{equation*}
p_{t_k}=i^kH_k(p,-p^\ast).
\end{equation*}
All the members of the AKNS and RAKNS hierarchies are covariant with respect to  space and time translations, generalized Galilean and scaling transformations (see \cite{Sm15a}).
 All members  of AKNS and RAKNS  hierarchies have a well known  common feature: for any integer $k$ there exist functions
\begin{equation*}
p(x,t_1,\ldots,t_k),
\end{equation*}
satisfying all equations of the hierarchy simultaneously.

Another integrable equations can be obtained using the function $p(x,t_1,\ldots,t_k)$ and substituting the variable $t$ directly into several phases $t_k$ simultaneously. For example, the integrable Hirota equation \cite{Hir73,Hir06,AAS2,Hir13,HeLiP} has the form

\begin{equation*}
ip_t+\alpha H_1(p,-p^\ast)-i\beta H_2(p,-p^\ast)=0. 
\end{equation*}
It is easy to see that this equation has a solution in the form $p(x,\alpha t,-\beta t,\ldots,t_k)$, where $p(x,t_1,\ldots,t_k)$ is an arbitrary solution of the equations of the AKNS hierarchy.
The more complex model, used in \cite{GNLS13, AA14}, is described by the equation
\begin{equation*}
ip_t+\alpha H_1(p,-p^\ast)-i\beta H_2(p,-p^\ast)+\gamma_1 H_3(p,-p^\ast)=0. 
\end{equation*}
It is easy to understand that it has solutions in the form $p(x,\alpha t,-\beta t,-\gamma_1 t,\ldots,t_k)$.
Next in order  equations are:
\begin{equation*}
ip_t+\alpha H_1(p,-p^\ast)-i\beta H_2(p,-p^\ast)+\gamma_1 H_3(p,-p^\ast)-i\gamma_2 H_4(p,-p^\ast)=0
\label{eq:hnls4}
\end{equation*}
and
\begin{equation*}
ip_t+\alpha H_1(p,-p^\ast)-i\beta H_2(p,-p^\ast)+\gamma_1 H_3(p,-p^\ast)-i\gamma_2 H_4(p,-p^\ast)+\gamma_3 H_5(p,-p^\ast)=0.
\end{equation*}
%

 The coefficients of the spectral curves equations do not depend on  times $t_k$. Therefore, the spectral curves equations depend only on the ``parent'' solution $p(x,t_1,\ldots,t_k)$: it does not depend on the choice  of the selected member of  hierarchy.

\section{Appel  equation and   spectral curves}

The  equation $\Psi_x=U \Psi$  has the following scalar form
\begin{equation} \label{1}
\begin{aligned}
&\psi_x=-i\lambda \psi+ip\phi,\\
&\phi_x=i\lambda \phi-iq\psi,
\end{aligned}
\end{equation}
%

%
%
or
\begin{equation} \label{eq:2a}
\psi_{xx}-\dfrac{p_x}{p}\psi_x+\left(\lambda^2-i\lambda\dfrac{p_x}{p}-pq\right)\psi=0.
\end{equation}

Let $\psi_1$ and $\psi_2$ are two linearly independent solutions of the equation
\begin{equation}
\psi_{xx}+P(x)\psi_x+Q(x)\psi=0. \label{eq:2}
\end{equation}
Then the function  $Y=\psi_1\psi_2$ satisfies Appel equation (\cite{WWe}, Part II, Chapter 14, Example 10; \cite{AP})
\begin{equation} \label{eq:AP}
Y_{xxx}+3PY_{xx}+(P'+4Q+2P^2)Y_x+(2Q'+4PQ)Y=0.
\end{equation}
Taking the coefficients  of \eqref{eq:2}  as in \eqref{eq:2a}  from  Appel equation we get:
\begin{multline} \label{3}
Y_{xxx}-3\dfrac{p_x}{p}Y_{xx}+\left(4\lambda^2-4i\lambda\dfrac{p_x}{p}+\dfrac{3p_x^2-p_{xx}p}{p^2}-4pq\right)Y_x-\\
-\left(4\lambda^2\dfrac{p_x}{p}+
2i\lambda\dfrac{p_{xx}p-3p_x^2}{p^2}+2pq_x-2qp_x\right)Y.
\end{multline}

Assume, that  \eqref{3} admits the solution of the form\footnote{Recall that in the case $P=0 $, $Q=q(x) - \lambda$, the  existence of solution polynomial in $\lambda$ of Appel equation  allows to isolate a very wide class of integrable potentials, including all finite-gap periodic and almost periodic potentials, all reflectionless potentials and more generally all Bargmann potentials}
\begin{equation} \label{poly}
Y=\sum_{j=0}^{g}\gamma_j(x)\lambda^{g-j}.
\end{equation}
 Substituting  \eqref{poly} into  \eqref{3} and equating the coefficients at the same powers of $\lambda$, we obtain  for  the coefficients $\gamma_j$  . The first two equations have the form
\begin{align*}
&4\gamma_0'-4\dfrac{p_x}{p}\gamma_0=0,\\
&4\gamma_1'-4\dfrac{p_x}{p}\gamma_1-4i\dfrac{p_x}{p}\gamma_0'
-2i\dfrac{p_{xx}p-3p_x^2}{p^2}\gamma_0=0.
\end{align*}
From these two equations we get
\begin{align*}
&\gamma_0(x)=c_0 p(x), 
\\
&\gamma_1(x)=\dfrac{i}2(c_0p_x+c_1 p). 
\end{align*}
Next equations have the form
\begin{multline} \label{eq:3->}
4\gamma_{j+2}'-4\dfrac{p_x}{p}\gamma_{j+2}-4i\dfrac{p_x}{p}\gamma_{j+1}'
-2i\dfrac{p_{xx}p-3p_x^2}{p^2}\gamma_{j+1}+\\
+\gamma_{j}'''-3\dfrac{p_x}{p}\gamma_{j}''-\left(4pq+\dfrac{p_{xx}p-3p_x^2}{p^2}\right)\gamma_j'-2(pq_x-qp_x)\gamma_j=0.
\end{multline}
Equations \eqref{eq:3->} allow to find the remaining coefficients using the recursive relations:
\begin{multline} \label{gamma:j+2}
\gamma_{j+2}=c_{j+2}p+p\int\left(i\dfrac{p_x}{p^2}\gamma'_{j+1}+i\dfrac{p_{xx}p-3p_x^2}{2p^3}\gamma_{j+1}-\right.\\
\left.-\dfrac1{4p}\gamma'''_{j}+\dfrac{3p_x}{4p^2}\gamma''_j+\left(\dfrac{p_{xx}p-3p_x^2}{4p^3}+q\right)\gamma'_j
-\dfrac{p_xq-q_xp}{2p}\gamma_j\right)dx.
\end{multline}
Assuming that  $c_0=1$, we obtain from \eqref{gamma:j+2} the following equalities
\begin{align*}
&\gamma_2=-\dfrac14(p_{xx}-2p^2q+c_1p_x+c_2p), 
\\
&\gamma_3=-\dfrac{i}8(p_{xxx}-6pqp_x+c_1(p_{xx}-2p^2q)+c_2p_x+c_3p). 
\end{align*}
Of course, since $\gamma_j\equiv0$ for $j>g$, equations \eqref{eq:3->} and \eqref{gamma:j+2} can only be used if $j\leq g-2$.
It can be shown that for $\gamma_j $ when   $ j\geq3$   we get:
\begin{equation*}
\gamma_j=\left(\dfrac{i}{2}\right)^j\left(H_{j-1}(p,q)+\sum_{k=1}^{j-2}c_kH_{j-1-k}(p,q)+c_{j-1}p_x+c_jp\right).
\end{equation*}
For $j=g-1$ and $j=g$ equation \eqref{eq:3->} takes the form
\begin{multline} \label{eq:g-1}
-4i\dfrac{p_x}{p}\gamma_{g}'-2i\dfrac{p_{xx}p-3p_x^2}{p^2}\gamma_{g}+\gamma_{g-1}'''-3\dfrac{p_x}{p}\gamma_{g-1}''-\\
-\left(4pq+\dfrac{p_{xx}p-3p_x^2}{p^2}\right)\gamma_{g-1}'-2(pq_x-qp_x)\gamma_{g-1}=0
\end{multline}
and
\begin{equation} \label{eq:g}
\gamma_{g}'''-3\dfrac{p_x}{p}\gamma_{g}''-\left(4pq+\dfrac{p_{xx}p-3p_x^2}{p^2}\right)\gamma_g'-2(pq_x-qp_x)\gamma_g=0.
\end{equation}
Knowing $p$ and $q$, we can find from equations \eqref{eq:g-1}, \eqref{eq:g} the values of the constants $c_k$.

Since the Wronskian $W[ \psi_1,\psi_2] :=(\psi_2)_x\psi_1 - (\psi_1)_x\psi_2 $  of any two solutions of  \eqref{eq:2}
 satisfy the differential equation  $W_x =-P(x)W $,  and, in \eqref{eq:2a}  $P(x) = - p_x\,p^{-1}(x)$,
we get
\begin{equation*}
W[\psi_1,\psi_2] = -2i\nu(\lambda)p(x),
\end{equation*}
where $\nu(\lambda) $  is $x$-independent function of $\lambda$.
Knowing the product of solutions $Y:=\psi_1\,\psi_2$ and their Wronskian $W$, we obtain
\begin{equation*}
\frac{\psi^{'}_2}{\psi_2}  =\frac{ Y' +W}{2Y}, \quad \frac{\psi^{'}_1}{\psi_1}  =\frac{ Y' - W}{2Y}.
\end{equation*}
Hence
\begin{equation} \label{psi}
\psi_{1,2}=\sqrt{Y}\exp\left(\pm i\nu(\lambda)\int\dfrac{p(x)dx}{Y(x)}\right).
\end{equation}

Substituting \eqref{psi} in \eqref{eq:2a} and simplifying, we obtain an equation of the spectral curve
\begin{equation} \label{curve}
\nu^2(\lambda)=\dfrac{Y^2}{p^2}\lambda^2-i\dfrac{p_xY^2}{p^3}\lambda
-\dfrac{4p^2qY^2-2 pY_{xx}Y+pY_x^2+2 p_xY_xY}{4p^3}.
\end{equation}

The right-hand side of the equation \eqref{curve} is a polynomial of degree $2g+2$. Its  coefficients are integrals of  motion for the  NLS equation.
These integrals can be found by substituting \eqref{poly} in  \eqref{curve} and simplifying.

\section{Examples}

In this section, we will assume that  $\im\; a=0$, $\im\; b=0$.

\subsection{Plane, solitary and ``dnoidale'' waves solutions}

Let us consider  solution of the nonlinear Schr\"odinger equation \eqref{eq:nls} in the form of a plane wave
\begin{equation*}
p(x,t)=ae^{2i(a^2-2b^2)t-2ibx}.
\end{equation*}
The spectral curve of this solution $\Gamma_0 $ has  both topological  and arithmetic genus $g=0= g_a$ :
\begin{equation*}
\Gamma_0 =\{\,(\nu,\lambda) : \nu^2=(\lambda-b)^2+a^2 \,\}.
\end{equation*}

The well known soliton solution  of the NLS equation  is described by the formula
\begin{equation*}
p(x,t)=\dfrac{2ae^{4i(a^2-b^2)t-2ibx}}{\cosh(2ax+8abt)}.
\end{equation*}
The spectral curve $\Gamma_{1s}$ of this solution is a degenerate (singular) elliptic curve of topological genus $g=0$ and  arithmetic genus  $g_a=1$
Here we have $c_1=4ib$.

\begin{equation} \label{curve.sol}
\Gamma_{1s} = \left\{\,(\nu,\lambda) :  \nu^2=\left((\lambda-b)^2+a^2\right)^2\right\}.
\end{equation}

The one-phase solution of the nonlinear Schr\"odinger equation with periodic amplitude has the form of a “dnoidal” wave:
\begin{equation*}
p(x,t)=2ae^{4i(2a^2-k^2a^2-b^2)t-2ibx}\dn(2ax+8abt;k),
\end{equation*}
where $\dn(x;k)$ is an elliptic Jacobi function \cite{Akhe}. The spectral curve of the ``dnoidal'' wave also has  both topological and arithmetic genus $1$ i.e. $g=1=g_a$ , and
($c_1=4ib$).
Contrary to the previous example  its spectral curve $\Gamma_{1dn}$ is  a \textbf{nonsingular elliptic curve }  i.e. it has only  simple branch points:
\begin{equation} \label{curve.dn}
\Gamma_{1dn} =\left \{ \,  (\nu,\lambda) :\nu^2=\left((\lambda-b)^2+a^2(1-k_1)^2\right)\left((\lambda-b)^2+a^2(1+k_1)^2\right) \right\},
\end{equation}
where $k^2+k_1^2=1$.

Thus, a plane wave is the null-phase solution of the nonlinear Schr\"odinger equation, and the solitary and ``dnoidal'' waves are one-phase. The phase of solitary and ``dnoidal'' waves are determined by equality
\begin{equation*}
X=2ax+8abt.
\end{equation*}
It is easy to see that in the limit at $k_1\to0$, $k\to1$ the spectral curve of the ``dnoidal'' wave goes into the spectral curve of a solitary wave, and the ``dnoidal'' wave goes into a solitary wave.

\subsection{Peregrine soliton}

Let us consider the well-known Peregrine soliton \cite{Per83}
\begin{equation*}
p(x,t)=\left(1-\dfrac{4(1+iT)}{X^2+T^2+1}\right)e^{2it},\quad X\equiv2x,\quad T\equiv 4t.
\end{equation*}
Performing the Galilean and scaling transformations  \cite{Sm15a}, we obtain the general form of the Peregrine soliton
\begin{equation}
\begin{gathered}
p(x,t)=a\left(1-\dfrac{4(1+iT)}{X^2+T^2+1}\right)e^{2i(a^2-2b^2)t-2ibx},\\ X\equiv2ax+8abt,\quad T\equiv 4a^2t.
\end{gathered}\label{eq:Per}
\end{equation}
Of course  the function \eqref{eq:Per} satisfies the NLS equation.

Substituting \eqref{eq:Per} and $q=-p^\ast$ into \eqref{eq:g-1} for $g=2$, we get
\begin{equation*}
c_1=6ib,\quad c_2=-6a^2-12b^2.
\end{equation*}
The function \eqref{eq:Per} satisfies equation \eqref{eq:g} for $g=2$. Therefore the function \eqref{eq:Per} is  a degenerate two-gap solution of the nonlinear Shr\"odinger equation. Calculating the spectral curve $\Gamma_P $  of the Peregrine soliton \eqref{eq:Per}, we get
\begin{equation*}
\Gamma_P := \left\{\, (\nu,\lambda) : \,\nu^2=\left((\lambda-b)^2+a^2\right)^3 \,\right\}
\end{equation*}
or
\begin{equation*}
\Gamma_P := \left\{\, (\nu,\lambda) : \,  \nu^2=(\lambda-b-ia)^3(\lambda-b+ia)^3\,\right\}.
\end{equation*}
Therefore a solution \eqref{eq:Per} with $a=\im(\lambda_1)$, $b=\re(\lambda_1)$ corresponds to a degenerated spectral curve $\Gamma_{Per}$:
\begin{equation*}
  \Gamma_{Per}:= \left\{ \, (\nu , \lambda) :  \nu^2=(\lambda-\lambda_1)^3(\lambda-\lambda_1^\ast)^3\,\right\}.
\end{equation*}

In the case of the canonical form of the Peregrine soliton, i.e. for $X=2x$, $T=4t$, the constants $c_k$ are equal to
\begin{equation*}
c_1=0,\quad c_2=-6,
\end{equation*}
and  the spectral curve becomes $\Gamma_p$:
\begin{equation} \label{curve.1}
\Gamma_p := \left\{\, (\nu, \lambda) : \,\nu^2=(\lambda^2+1)^3\,\right\}.
\end{equation}
It is clear that all spectral curves connected with Peregrin soliton are singular curves of arithmetic genus $g_a=2$ and of topological genus $g=0$.

\subsection{The Kuznetsov-Ma soliton and the Akhmediev breather}

The Kuznetsova-Ma soliton \cite{Ku77e,Ma79} is a two-phase solution periodic in $x$. Let us write it in the form  \cite{Sm16tmfe}
\begin{equation} \label{eq:kuma}
p(x,t)=\left(1-\dfrac{2k(k\cosh(k\vk T)+i\vk\sinh(k\vk T))}
{\sqrt{\vk^2+k^2}\cosh(k\vk T)-\vk\cos(k X)}\right)e^{2i(\vk^2+k^2)t},
\end{equation}
where $X=2x$, $T=4t$, $k=\sin\theta$, $\vk=\cos\theta$ ($\theta$ is a parameter of the solution).

From \eqref{eq:g-1} and \eqref{eq:g} for $g=2$ we get
\begin{equation*}
c_1=0,\quad c_2=-2-4\cos^2\theta.
\end{equation*}
The spectral curve of the Kuznetsov-Ma soliton \eqref{eq:kuma} is given by the equation
\begin{equation*}
\nu^2=(\lambda^2+1)(\lambda^2+\cos^2\theta)^2.
\end{equation*}
For the case of the Kuznetsov-Ma soliton \eqref{eq:kuma} with arguments  $X\equiv2ax+8abt$, $T\equiv 4a^2t$ the constant $c_1$ and $c_2$ are equal
\begin{equation*}
c_1=6ib,\quad c_2=-12b^2-(2+4\cos^2\theta)a^2,
\end{equation*}
and the spectral curve is given by the equation
\begin{equation*}
\nu^2=(\lambda^2-2b\lambda+b^2+a^2)(\lambda^2-2b\lambda+b^2+a^2\cos^2\theta)^2
\end{equation*}
or
\begin{equation} \label{curve.KuMa}
\nu^2=\left((\lambda-b)^2+a^2\right)\left((\lambda-b)^2+a^2\cos^2\theta\right)^2.
\end{equation}

The Akhmediev breather \cite{Ak85e} is a two-phase solution periodic in $t$. It can be obtained from the Kuznetsov-Ma soliton  \eqref{eq:kuma} by substitution $\theta\to i\theta$ \cite{Sm16tmfe}:
\begin{equation} \label{eq:akh}
p(x,t)=\left(1+\dfrac{2k(k\cos(k\vk T)+i\vk\sin(k\vk T))}
{\sqrt{\vk^2-k^2}\cos(k\vk T)-\vk\cosh(kX)}\right)e^{2i(\vk^2-k^2)t},
\end{equation}
where $X=2x$, $T=4t$, $k=\sinh\theta$, $\vk=\cosh\theta$.
Correspondingly, the constants $c_1$ and $c_2$ are equal
\begin{equation*}
c_1=0,\quad c_2=-2-4\cosh^2\theta,
\end{equation*}
and the spectral curve of Akhmediev breather \eqref{eq:akh} is given by equation
\begin{equation*}
\nu^2=(\lambda^2+1)(\lambda^2+\cosh^2\theta)^2.
\end{equation*}
Let us remark that for $X\equiv2ax+8abt$, $T\equiv 4a^2t$ the constants $c_1$ and $c_2$ are equal
\begin{equation} \label{curve.Akh}
c_1=6ib,\quad c_2=-12b^2-(2+4\cosh^2\theta)a^2,
\end{equation}
and the spectral curve is given by equation
\begin{equation*}
\nu^2=(\lambda^2-2b\lambda+b^2+a^2)(\lambda^2-2b\lambda+b^2+a^2\cosh^2\theta)^2
\end{equation*}
or
\begin{equation*}
\nu^2=\left((\lambda-b)^2+a^2\right)\left((\lambda-b)^2+a^2\cosh^2\theta\right)^2.
\end{equation*}
For all exemples of this subsection the related spectral curves are singular algebraic curves of arithmetic genus $g_a= 2$ and of topological genus $g=0$. The spectral curves considered here have a couple of simple branch points and a couple  of  double branch points each.

\subsection{Rank-2 rogue wave solution }

Rank 2  rogue waves solution reads \cite{Sm16tmfe}
\begin{equation} \label{u2X}
\Psi_2(X,T_1,T_2,T_3):=\left(1-12\dfrac{G(X,T_1,T_2,T_3) +iH(X,T_1,T_2,T_3)}{Q(X,T_1,T_2,T_3)}\right)e^{2it_1-6it_3+20it_5+\ldots},
\end{equation}
where
\begin{align*}
G(X,T_1,T_2,T_3)=&(X^2+3T_1^2+3)^2-4T_1^4+2XT_2+2T_1T_3-12,\\
H(X,T_1,T_2,T_3)=&T_1(X^2+T_1^2+1)^2+2XT_1T_2 +T_3(T_1^2-X^2-1)-8T_1(X^2+2),\\
Q(X,T_1,T_2,T_3)=&(X^2+T_1^2+1)^3+T_2^2+2XT_2(3T_1^2-X^2+3) +T_3^2 +\\
&+2T_1T_3(T_1^2-3X^2+9)+24T_1^4-24T_1^2X^2+96T_1^2+24X^2+8.
\end{align*}
Here
\begin{align*}
&X=2x-12t_2+60t_4+\ldots,\\
&T_1=4t_1-24t_3+120t_5+\ldots,\\
&T_2=-48t_2+480t_4+\ldots,\\
&T_3=-96t_3+960t_5+\ldots
\end{align*}

It is easy to see, that this solution is four-phase. First phase is $X$, \dots, fourth phase is $T_3$. Hence,  for arithmetic genus $g_a$ of the related spectral curve we have $g_a=4$.
Calculating the constants $c_k$, we get
\begin{equation*}
c_4=-6c_2-30,\quad c_3=-6c_1,\quad c_2=-10,\quad  c_1=0.
\end{equation*}
The spectral curve $\Gamma_2 $ for the solution \eqref{u2X} is
\begin{equation} \label{curve.2}
\Gamma_2 := \left\{ \,(\nu, \lambda\,): \nu^2=(\lambda^2+1)^5  \right\}
\end{equation}

\subsection{Rank-3  rogue wave  and its spectral curve}

A ``freak wave'' of rank 3 is defined by the following equalities

\begin{equation} \label{sol:3}
\Psi_3(X,T_1,\ldots,T_5)=\left(1-24\dfrac{G(X,T_1,\ldots,T_5)+iH(X,T_1,\ldots,T_5)} {Q(X,T_1,\ldots,T_5)}\right)e^{2it_1-6it_3+20it_5+\ldots},
\end{equation}
where
\begin{align*}
&G(X,T_1,\ldots,T_5)=X^{10}+\sum_{j=0}^8g_jX^j,\\
&g_8=15T_1^2+15,\quad g_7=0,\\
&g_6=50T_1^4-60T_1^2-80T_1T_3+210,\\
&g_5=120T_1^2T_2+120T_2+18T_4,\\
&g_4=70T_1^6-150T_1^4-200T_1^3T_3+450T_1^2-(600T_3-30T_5)T_1+150T_2^2-50T_3^2-450,\\
&g_3=400T_1^4T_2+(2400T_2+60T_4)T_1^2+800T_1T_2T_3-1200T_2+60T_4,\\
&g_2=45T_1^8+420T_1^6+6750T_1^4+(2400T_3+180T_5)T_1^3-{}\\
&\quad{}-(300T_2^2-900T_3^2+13500)T_1^2-(7200T_3-180T_5)T_1-300T_2^2-300T_3^2-675,\\
&g_1=280T_1^6T_2-(600T_2+150T_4)T_1^4-800T_1^3T_2T_3+(1800T_2-540T_4)T_1^2-\\
&\quad{}-(2400T_2T_3-120T_2T_5+120T_3T_4)T_1-200T_2^3-200T_2T_3^2-1800T_2+90T_4,\\
&g_0=11T_1^{10}+495T_1^8+120T_1^7T_3+2190T_1^6+(2040T_3-42T_5)T_1^5+(350T_2^2+150T_3^2-7650)T_1^4+\\
&\quad{}+(1800T_3-420T_5)T_1^3+(300T_2^2-120T_2T_4+300T_3^2-120T_3T_5-2025)T_1^2-\\
&\quad{}-(200T_2^2T_3+200T_3^3-1800T_3+90T_5)T_1+750T_2^2-120T_2T_4+2550T_3^2-\\
&\quad{}-240T_3T_5+6T_4^2+6T_5^2+675;
\end{align*}
\begin{align*}
&H(X,T_1,\ldots,T_5)=T_1X^{10}+\sum_{j=0}^8h_jX^j,\\
&h_8=5T_1^3-15T_1-5T_3,\quad h_7=0,\\
&h_6=10T_1^5-140T_1^3-40T_1^2T_3-150T_1+40T_3-5T_5,\\
&h_5=40T_1^3T_2-6(20T_2-3T_4)T_1-40T_2T_3,\\
&h_4=10T_1^7-210T_1^5-50T_1^4T_3-450T_1^3-15(20T_3-T_5)T_1^2+50(3T_2^2-T_3^2-27)T_1+\\
&\quad{}+15(10T_3-T_5),\\
&h_3=80T_1^5T_2+20(40T_2+T_4)T_1^3+400T_1^2T_2T_3-60(20T_2+T_4)T_1-20(20T_2T_3-T_2T_5+T_3T_4),\\
&h_2=5T_1^9-60T_1^7+1710T_1^5+15(80T_3+3T_5)T_1^4-100(T_2^2-3T_3^2+63)T_1^3-90T_1^2T_5+\\
&\quad{}+75(4T_2^2+4T_3^2+63)T_1+5(20T_2^2T_3+20T_3^3+720T_3-27T_5),\\
&h_1=40T_1^7T_2-30(28T_2+T_4)T_1^5-200T_1^4T_2T_3-60(30T_2+T_4)T_1^3-60(20T_2T_3-T_2T_5+T_3T_4)T_1^2-\\
&\quad{}-50(4T_2^3+4T_2T_3^2+108T_2-9T_4)T_1+60(10T_2T_3-T_2T_5+T_3T_4),\\
&h_0=T_1^{11}+25T_1^9+15T_1^8T_3-870T_1^7+(100T_3-7T_5)T_1^6+10(7T_2^2+3T_3^2-963)T_1^5-\\
&\quad{}-75(58T_3+T_5)T_1^4-5(100T_2^2+8T_2T_4+100T_3^2+8T_3T_5+495)T_1^3-\\
&\quad{}-5(20T_2^2T_3+20T_3^3+1980T_3-99T_5)T_1^2-3(350T_2^2-40T_2T_4+550T_3^2-2T_4^2-2T_5^2-1575)T_1+\\
&\quad{}+5(20T_2^2T_3-4T_2^2T_5+8T_2T_3T_4-60T_3^3+4T_3^2T_5+315T_3-9T_5);
\end{align*}
\begin{align*}
&Q(X,T_1,\ldots,T_5)=(X^2+T_1^2+1)^6-20T_2X^9+\sum_{j=0}^8q_jX^j,\\
&q_8=-120T_1^2-60T_1T_3+120,\quad q_7=-12T_4,\\
&q_6=-240T_1^4-160T_1^3T_3+480T_1^2+(960T_3-60T_5)T_1+60T_2^2+140T_3^2+2320,\\
&q_5=120T_1^4T_2-(720T_2-108T_4)T_1^2-480T_1T_2T_3+1080T_2+108T_4,\\
&q_4=-120T_1^5T_3-1440T_1^4-(3600T_3-60T_5)T_1^3+(900T_2^2-300T_3^2+13440)T_1^2-\\
&\quad{}-(5400T_3-540T_5)T_1+900T_2^2+120T_2T_4-1500T_3^2+120T_3T_5+3360,\\
&q_3=160T_1^6T_2+(7200T_2+60T_4)T_1^4+1600T_1^3T_2T_3+(21600T_2-360T_4)T_1^2+\\
&\quad{}+(4800T_2T_3+240T_2T_5-240T_3T_4)T_1+400T_2^3+400T_2T_3^2-7200T_2+540T_4,\\
&q_2=240T_1^8+13440T_1^6+(4320T_3+108T_5)T_1^5-(300T_2^2-900T_3^2-78240)T_1^4+\\
&\quad{}+(43200T_3+1080T_5)T_1^3+(1800T_2^2+16200T_3^2-36480)T_1^2+\\
&\quad{}+(1200T_2^2T_3+1200T_3^3-64800T_3+2700T_5)T_1-2700T_2^2+900T_3^2-720T_3T_5+\\
&\quad{}+36T_4^2+36T_5^2+12144,\\
&q_1=60T_1^8T_2+(240T_2-60T_4)T_1^6-480T_1^5T_2T_3-(5400T_2+1620T_4)T_1^4-\\
&\quad{}-(14400T_2T_3-240T_2T_5+240T_3T_4)T_1^3-(1200T_2^3+1200T_2T_3^2-54000T_2+5940T_4)T_1^2-\\
&\quad{}-(21600T_2T_3-2160T_2T_5+2160T_3T_4)T_1-1200T_2^3+240T_2^2T_4-6000T_2T_3^2+\\
&\quad{}+480T_2T_3T_5-240T_4T_3^2+13500T_2-540T_4,\\
&q_0=120T_1^{10}+20T_1^9T_3+3720T_1^8+(1200T_3-12T_5)T_1^7+(140T_2^2+60T_3^2+15280)T_1^6+\\
&\quad{}+(5400T_3-612T_5)T_1^5+(900T_2^2-120T_2T_4-1500T_3^2-120T_3T_5+143760)T_1^4-\\
&\quad{}-(400T_2^2T_3+400T_3^3-82800T_3-540T_5)T_1^3+\\
&\quad{}+(8100T_2^2+720T_2T_4+18900T_3^2+36T_4^2+36T_5^2+93144)T_1^2+\\
&\quad{}+(6000T_2^2T_3-240T_2^2T_5+480T_2T_3T_4+1200T_3^3+240T_3^2T_5+83700T_3-2700T_5)T_1+400T_2^4+\\
&\quad{}+800T_3^2T_2^2+400T_3^4+9900T_2^2-1080T_2T_4+24300T_3^2-1800T_3T_5+36T_4^2+36T_5^2+2024.
\end{align*}

Here
\begin{align*}
&X=2x-12t_2+60t_4+\ldots,\\
&T_1=4t_1-24t_3+120t_5\ldots,\\
&T_2=-48t_2+480t_4+\ldots,\\
&T_3=-96t_3+960t_5+\ldots,\\
&T_4=-3840t_4+\ldots,\\
&T_5=-7680t_5+\ldots
\end{align*}

The rogue wave solution of the  rank 3 is a six-phase solution of the AKNS hierarchy equations of the aritmetic genus $g_a =6$. Calculating the constants $c_k$, we get
\begin{gather*}
c_6=-6c_4-30c_2-140,\quad
c_5=-6c_3-30c_1, \quad
c_4=-10c_2-70,\\
c_3=-10c_1,\quad
c_2=-14,\quad
c_1=0.
\end{gather*}
It folllows  from   \eqref{curve}  that the spectral curve $\Gamma_3$, corresponding to  solution  \eqref{sol:3} is
\begin{equation} \label{curve.3}
 \Gamma_3 = \{\, (\nu, \lambda) \,:\, \nu^2 = (\lambda^2+1)^7 \},
 \end{equation}
 i.e. it represents a singular algebraic curve of the arithmetic  genus 6.

\section*{Concluding remarks}


\begin{itemize}
\item The spectral curves $\Gamma_{N}$, corresponding to  Matveev-Dubard-Smirnov   \cite{DubMat11, DubMat13, Sm16tmfe}   quasi-rational rank $N$ solutions  of  AKNS hierarchy
equations are:
\begin{equation*}
\Gamma{_N}:= \left\{(\,\nu, \lambda \,) \,: \,\nu^2=(\lambda^2+1)^{2N+1}\,\right\},
\end{equation*}
i.e. it represents  singular algebraic curve of the arithmetic genus $2N$ with $2 $ branch points:  $(0,i)$  and $ (0,-i)$ of mutiplicity $2N+1$  each.
\item Polynomials in the RHS of  the spectral curves related with  solutions, containing trigonometric or hyperbolic functions, have one pair of simple complex conjugate roots and double complex conjugate roots. Solutions of this type can be obtained by the Darboux transformation of a plane wave.
\item  Polynomials in the RHS of the spectral curves equations  of multi-solitons solutions have only double complex conjugate roots. Multi-solitons solutions can be obtained by the Darboux transformation of the zero seed solution. They can also be obtained by passage to the limit in theta functional formulas related with nonsingular hyperelliptic spectral curves.
\item  it will be interesting to investigate   the solutions  of  whole AKNS hierarchy equation for the spectral curve $\Gamma_r$  defined by
$$
\Gamma_r =  \left\{\, (\nu, \lambda )\, :\, \nu^2 =(\lambda^2 -1) (\lambda^2 - k^{-2}) \prod_{j=1}^{N}\,(\lambda^2 -  \lambda^{2}_{j})^{2}\, \right\}; $$
$$  1 < \lambda_1 <\lambda_2  < \ldots  < \lambda_N  < k^{-1}.$$
For  the  NLS equation it was done  in \cite{IAKe} and  in the section 4.5  of \cite{BBEIM} in   the context of studying multiphase modulations of the  dnoidal wave solution.
The related spectral curve obviously has a topological genus $g=1$ and the arithmetic genus $g_a = 2N + 1$ . It has $4 $ simple branch points and  $ 2N $ double branch points.
%
%
%
%
\end{itemize}

\section*{Aknowledgments}
This work was partially supported by RFBR   grant  16-01-00518. The first of the authors (AS) appreciates a kind hospitality and
financial support  at the International workshop  on Gromov-Witten Theory In Hefei, China, where the part of the results of this work was
 first time  reported,\\  (see http://siqiliu.com/conf/GWIH-2017/).



\end{document}